\documentclass[onecolumn]{pasj00}

\begin{document}
\SetRunningHead{T. Nishimichi et al.}
{Bispectrum and nonlinear biasing of galaxies}
\Received{2006/09/15}
\Accepted{2006/12/10}

\title{Bispectrum and Nonlinear Biasing of Galaxies: 
Perturbation Analysis, Numerical Simulation and SDSS Galaxy Clustering}
\author{%
Takahiro \textsc{Nishimichi}\altaffilmark{1},
Issha \textsc{Kayo}\altaffilmark{2},
Chiaki \textsc{Hikage}\altaffilmark{2},
Kazuhiro \textsc{Yahata}\altaffilmark{1},\\
Atsushi \textsc{Taruya}\altaffilmark{1},
Y. P. \textsc{Jing}\altaffilmark{3},
Ravi K. \textsc{Sheth}\altaffilmark{4}, and,
Yasushi \textsc{Suto}\altaffilmark{1}}
\altaffiltext{1}
{Department of Physics, School of Science,
The University of Tokyo, Tokyo 113-0033}
\altaffiltext{2}
{Department of Physics and Astrophysics,
Nagoya University, Chikusa, Nagoya 464-8602}
\altaffiltext{3}
	     {The Partner Group of MPI fur Astrophysik, 
Shanghai Astronomical Observatory,\\
Nandan Road 80, Shanghai 200030, China}
\altaffiltext{4}
{Department of Physics \& Astronomy, University 
of Pennsylvania, PA 19130, USA}
\email{nishimichi@utap.phys.s.u-tokyo.ac.jp}
\maketitle
\begin{abstract}
We consider nonlinear biasing models of galaxies with particular
attention to a correlation between linear and quadratic biasing
coefficients, $b_1$ and $b_2$. We first derive perturbative expressions
for $b_1$ and $b_2$ in halo and peak biasing models.  Then we compute

power spectra and bispectra of dark matter particles and halos using
N-body simulation data and of volume-limited subsamples of Sloan Digital
Sky Survey (SDSS) galaxies, and determine their $b_1$ and $b_2$. We find
that the values of those coefficients at linear regimes ($k<0.2h{\rm
Mpc}^{-1}$) are fairly insensitive to the redshift-space distortion and
the survey volume shape. The resulting normalized amplitudes of
bispectra, $Q$, for equilateral triangles, are insensitive to the values
of $b_1$ implying that $b_2$ indeed correlates with $b_1$.  The present
results explain the previous finding of Kayo et al. (2004) for the
hierarchical relation of three-point correlation functions of SDSS
galaxies.  While the relations between $b_1$ and $b_2$ are
quantitatively different for specific biasing models, their
approximately similar correlations indicate a fairly generic outcome of
the biasing due to the gravity in primordial Gaussian density fields.
\end{abstract}
\section{Introduction}

One of the major uncertainties in precise cosmology is the galaxy
biasing relative to the underlying mass distribution. It hampers
extracting the cosmological information from the large scale structure
of the universe. In particular, the biasing is sensitive to the unknown
physical conditions of galaxy formation. Thus its phenomenological
parametrization and understanding is crucial in advancing our knowledge
of the evolution of the universe.

A reasonable approximation often adopted is a local linear biasing model, 
which assumes a simple scale-independent relation between the density 
contrast fields of galaxy and mass: 
\begin{equation}
\delta_g({\bf x},z) = b(z)\delta({\bf x},z).
\label{eq:dg-d}
\end{equation}
In the above, the density contrast of the mass is defined as  
\begin{equation}
\delta({\bf x},z) \equiv \frac{\rho({\bf x},z)-\bar{\rho}(z)}{\bar{\rho}(z)},
\end{equation}
where the over-bar indicates the mean over the entire universe. The
number density contrast of galaxy density field, $\delta_g$, is defined
similarly. Because this model contains only a single time-dependent
parameter $b(z)$, it is not surprising that the model cannot describe
the observed features of galaxy clustering accurately.  Furthermore, the
three-point correlation function of SDSS galaxies \citep{Kayo2004}
indicates the importance of the higher-order correction to equation
(\ref{eq:dg-d}); while the galaxy two-point correlation functions are
well represented by the linear biasing model (e.g., \cite{Zehavi2005}),
the normalized amplitudes $Q$ of the corresponding three-point functions 
for equilateral triangles do not show the expected scaling with respect 
to $b(z)$:
\begin{equation}
Q\propto\frac{1}{b}.
\label{eq:Qinvb}
\end{equation}
In reality, however, $Q$ for equilateral triangles calculated from 
SDSS galaxies in redshift space proves to be
almost scale-independent, and follows the hierarchical relation
approximately, $Q=0.5\sim1.0$. Moreover its dependence on the
morphology, color, and luminosity is not statistically significant.
Given the robust morphological, color and luminosity dependences of the
two-point correlation function, \citet{Kayo2004} argued that galaxy
biasing is complex and requires a contrived relation between the linear
biasing and its higher order correction terms. Another possibility is
that the observed value is significantly contaminated by the
redshift-space distortion and does not properly reflect the actual
clustering information in real space. Indeed the previous N-body studies
of the redshift-space distortion on three- and four-point correlation
functions suggest that this is the case in nonlinear regimes
\citep{Suto93,MS94,SM94}.  The interpretation is further complicated by
the fact that the higher-order statistics is sensitive to relatively
rare large-scale structures in particular samples \citep{Nichol06}.

The main purpose of the present paper is to provide a physical
explanation for the approximate hierarchical relation for $Q$ of SDSS
galaxy clustering on the basis of perturbation analysis and numerical
simulations.  This is a step toward the understanding of the nonlinear
nature of biasing. Specifically we employ a local {\it nonlinear}
biasing model \citep{FG93}:
\begin{equation}
\delta_b({\bf x},z) = \sum_ {n=0}^\infty\frac{b_n(z)}{n!}
\left[\delta({\bf x},z)\right]^n.
\label{eq:db-d}
\end{equation}
We consider density peaks, dark matter halos, simulated halos, and the
SDSS galaxies as specific examples for the density fields $\delta_b$ in
the left hand side.

The outline of the paper is as follows; section \ref{sec:AM} briefly
summarizes the halo and peak biasing models as analytically tractable
examples. We compute the linear and quadratic biasing coefficients
perturbatively and find that these models roughly agree with the
observed hierarchical relation.  To be more realistic, we analyze
simulated halo catalogs and SDSS galaxies in section
\ref{sec:simulation}, and find that the perturbative result is valid
even if we take account of a variety of selection effects and
particularly redshift-space distortion.  The quantitative agreement
between the two indicates a generic correlation among the biasing
coefficients in the SDSS galaxies. Finally, section \ref{sec:SD} is
devoted to the summary and conclusion of the paper.

\section{Perturbative predictions in halo and peak biasing models}
\label{sec:AM}

\subsection{Basic statistics and perturbation biasing models}
Throughout the present analysis, we work in the Fourier space.  For
definiteness, we adopt the definition of the Fourier transform of an
arbitrary function $g({\bf x})$ as
\begin{equation}
g({\bf k}) \equiv \int \frac{d^3{\bf x}}{(2\pi)^3}g({\bf x})e^{-i{\bf k}
\cdot{\bf x}}.
\end{equation}
Then the power spectrum and the bispectrum of the mass density field 
are defined as 
\begin{eqnarray}
\langle\delta({\bf k}_1)\delta({\bf k}_2)\rangle &\equiv& 
P(k_1)\delta_D({\bf k}_1+{\bf k}_2),\\
\langle\delta({\bf k}_1)\delta({\bf k}_2)\delta({\bf k}_3)\rangle &\equiv& 
B(k_1,k_2,k_3)\delta_D({\bf k}_1+{\bf k}_2+{\bf k}_3),
\end{eqnarray}
where $\delta_D$ denotes the Dirac delta and we assume the universe is
isotropic and homogeneous. We introduce the normalized amplitude of the
mass bispectrum as
\begin{eqnarray}
Q_m(k_1,k_2,k_3) \equiv \frac{B(k_1,k_2,k_3)}{P(k_1)P(k_2)+P(k_2)P(k_3)+
P(k_3)P(k_1)}.
\label{eq:Qm}
\end{eqnarray}
Note that strictly speaking, the above statistic is different from that
defined in configuration space\footnote{In this paper, we use the term
``real space" in order to imply the analysis without redshift-space distortion
in k-space. Thus we reserve the term ``configuration space" in order to
distinguish from k-space.}. Since both are expected to be nearly
identical in a weakly nonlinear regime, we focus on the Fourier space 
analysis throughout this paper (see also \cite{Verde2002} 
and \cite{Hikage2005}). 
The analysis in configuration space is now in progress.

The above statistics can be generalized in a straightforward manner to
any biased field.  If one keeps the linear and quadratic terms in
equation (\ref{eq:db-d}), one obtains
\begin{eqnarray}
P_b(k) &=& b_1^2P(k),\label{eq:pb}\\
Q_b(k_1,k_2,k_3) &=& \frac{1}{b_1}
\left[Q_m(k_1,k_2,k_3)+\frac{b_2}{b_1}\right].\label{eq:Qb}
\end{eqnarray}
Therefore, $Q_b$ for the biased field is directly dependent on the
quadratic biasing coefficient $b_2$ unlike in the case of the power
spectrum. As we show below, this term plays an important role to
approximately satisfy the hierarchical relation.  In order to understand
the qualitative relations among the biasing coefficients, we first
consider analytically workable biasing models, i.e., halo and peak
biasing.

\subsection{Halo biasing}
\label{MJW}
We follow the formalism of \citet{Mo1996} for halo biasing.  We define
(unconditional) mass function, $n_{\rm halo}(m,z)$ of halos with mass
$m$ at redshift $z$. We rewrite the unconditional mass function as
\begin{equation}
\frac{m^2n_{\rm halo}(m,z)}{\bar{\rho}}\frac{dm}{m} 
= \nu f(\nu)\frac{d\nu}{\nu},
\hspace{3em}
\nu(m,z) \equiv \delta_{\rm sc}(z)/\sigma(m).
\end{equation}
We denote by $\sigma(m)$ the linearly extrapolated value of the rms of
the initial density fluctuation field smoothed with a tophat filter of
scale $R=(3m/4\pi\bar{\rho})^{1/3}$, where $\bar\rho$ is the mean
comoving background density.  The characteristic density contrast for
spherical collapse is given as
\begin{equation}
\delta_{\rm sc}(z) \simeq 
\frac{3(12\pi)^{2/3}}{20}[1+0.0123\log_{10}\Omega_m(z)]/D(z),
\end{equation}
where $\Omega_m(z)$ is the matter density in units of the critical
density at redshift $z$, and $D(z)$ is the linear growth rate
\citep{Kitayama1996}.  In the above equation and throughout the paper,
we assume a spatially-flat model with non-vanishing cosmological
constant.  Specifically, we adopt the present value of mass density
$\Omega_m=0.3$, the cosmological constant $\Omega_\Lambda=0.7$, the
Hubble constant $h=H_0/(100 {\rm km}~{\rm s}^{-1}{\rm Mpc}^{-1})=0.7$,
and the amplitude of the density fluctuations smoothed over $8h^{-1}{\rm
Mpc}$ $\sigma_8 = 0.9$.

There exist two popular models for $f(\nu)$.  One is based on a
spherical collapse model \citep{Press1974}:
\begin{eqnarray}
\nu f_{\rm PS}(\nu) &=& 2\sqrt{\frac{\nu^2}{2\pi}}\exp(-\nu^2/2).
\label{eq:PS}
\end{eqnarray}
The other is based on an ellipsoidal collapse model \citep{Sheth1999,Sheth2001}: 
\begin{equation}
\nu f_{\rm ST}(\nu) = 2A(p)\left(1+\frac{1}{(q\nu^2)^{p}}\right)
\sqrt{\frac{q\nu^2}{2\pi}}\exp(-q\nu^2/2),
\label{eq:ST}
\end{equation}
where $p\approx0.3$,
$A(p)=[1+2^{-p}\Gamma(1/2-p)/\sqrt{\pi}]^{-1}\approx0.3222$, and
$q\approx0.75$ \citep{Cooray2002}.

Consider a comoving volume $V$ and its total mass $M$ at redshift
$z_0$.  Let $n_{\rm halo}(m,z_1|M,V,z_0)dm$ denote the conditional
number density of halos whose mass is between $m$ and $m+dm$ at $z_1$. A
reasonable approximation for $n_{\rm halo}(m,z_1|M,V,z_0)$ is obtained
by applying the extended Press-Schechter theory
(e.g., \cite{Bower1991,BCEK1991}):
\begin{eqnarray}
\frac{m^2n_{\rm halo}(m,z_1|M,V,z_0)}{\bar{\rho}}\frac{dm}{m} 
&=& \nu_{10}f(\nu_{10})\frac{d\nu_{10}}{\nu_{10}},\\
\nu_{10} &=& 
\frac{\delta_{\rm sc}(z_1)-\delta_0(z_0)}{\sqrt{\sigma^2(m)-\sigma^2(M)}},
\end{eqnarray}
where $\delta_0(z_0)$ is the linearly extrapolated mass density contrast
at $z_0$.  If we define the actual density contrast $\delta=M/(\bar\rho
V)-1$, $\delta_0(z_0)$ is written as
\begin{eqnarray}
\frac{\delta_0}{1+z_0} &=& \sum_{i=0}^\infty a_i\delta^i,
\label{eq:D0D}\\
a_0 = 0; \hspace{0.5em} a_1 = 1; \hspace{0.5em} a_2 = -\frac{17}{21}; 
\hspace{0.5em} a_3 &=& \frac{341}{567}; 
\hspace{0.5em} a_4 = -\frac{55805}{130977}; \hspace{0.5em} ...,
\end{eqnarray}
where the above coefficients are obtained assuming the spherical
collapse model \citep{Bernardeau1994}.

Now one can write down the corresponding conditional density contrast of
halos:
\begin{equation}
\delta_{\rm halo}(m,z_1|M,V,z_0) 
= \frac{n_{\rm halo}(m,z_1|M,V,z_0)(1+\delta)}{n_{\rm halo}(m,z_1)}-1.
\label{eq:Dhalo-n}
\end{equation}
Expanding equation (\ref{eq:Dhalo-n}) in terms of $\delta_0$ in the
limit of $\sigma(M)\rightarrow 0$ yields the biasing coefficients of
halos \citep{MJW1997,Scoccimarro2001,Cooray2002}:
\begin{eqnarray}
b_1(m,z) &=& 1+\epsilon_1+E_1,\nonumber\\
b_2(m,z) &=& 2(1+a_2)(\epsilon_1+E_1)+\epsilon_2+E_2,
\label{eq:halo_bias}
\end{eqnarray}
where
\begin{eqnarray}
\epsilon_1 &=& \frac{q\nu^2(m,z)-1}{\delta_{\rm sc}(z)},\qquad
\epsilon_2 = \frac{q\nu^2(m,z)}{\delta_{\rm sc}(z)}
\left(\frac{q\nu^2(m,z)-3}{\delta_{\rm sc}(z)}\right),\\
E_1 &=& \frac{2p}{\delta_{\rm sc}(z)[1+(q\nu^2(m,z))^p]},\qquad
\frac{E_2}{E_1} = \frac{1+2p}{\delta_{\rm sc}(z)}+2\epsilon_1.
\end{eqnarray}
The above results for ellipsoidal collapse reduce to those for
spherical collapse, if one sets $p=0$ and $q=1$.

\subsection{Peak biasing}

The biasing coefficients of the peak model can be obtained similarly as
the halo model described in the above subsection. The conditional and
unconditional number densities of peaks with peak height $\nu \equiv
\delta/\sigma$ are derived in \citet{BBKS1986}. In this case, $\sigma$
denotes the rms value of density fluctuation smoothed over
$R\equiv(3m/4\pi\bar\rho)^{1/3}$, and $\delta$ is the density contrast
of the peaks in the smoothed field.  Substituting those formulae into
the right-hand side of equation (\ref{eq:Dhalo-n}), one similarly
obtains the biasing coefficients for the peak model \citep{MJW1997}:
\begin{eqnarray}
b_1(\nu,z) &=& 1+\frac{\nu^2+g_1}{\delta_{\rm sc}(z)},\\
b_2(\nu,z) &=& 2(1+a_2)\frac{\nu^2+g_1}{\delta_{\rm sc}(z)}+
\left(\frac{\nu}{\delta_{\rm sc}(z)}\right)^2
\left(\nu^2-1+2g_1+\frac{2g_2}{\nu^2}\right),
\end{eqnarray}
where the above functions, $g_1$ and $g_2$, are defined in equation (25)
of \citet{MJW1997}.
\subsection{Results of Analytic Models}
We consider the above three models, spherical halo, ellipsoidal halo,
and peak, to explore the correlation between $b_1$ and $b_2$
analytically.
\begin{figure}[!t]
   \centering \FigureFile(160mm,80mm){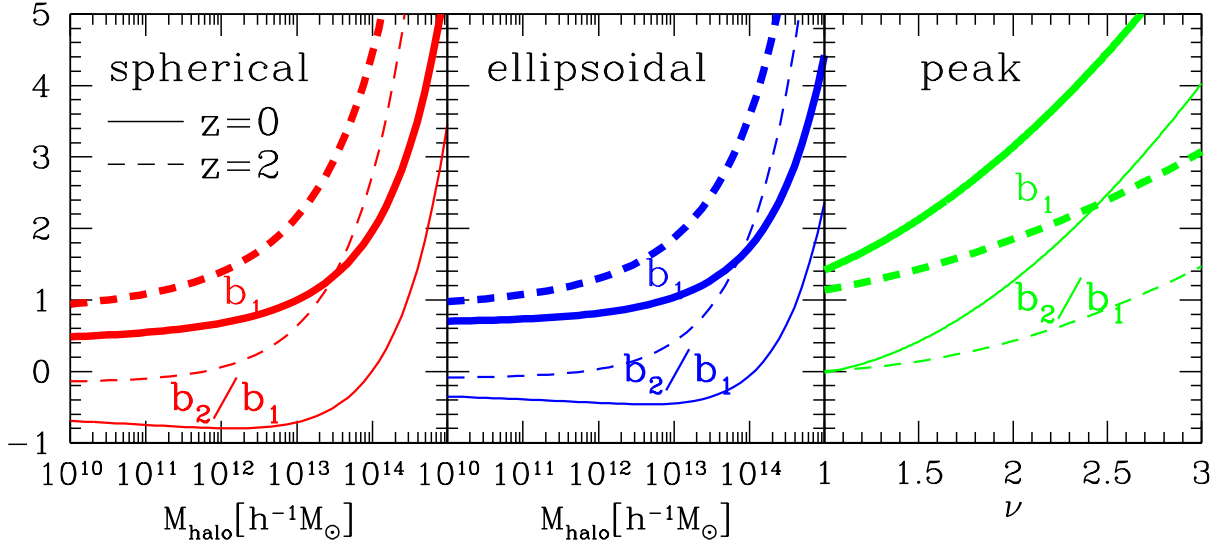} 
\caption{ biasing coefficients in the three perturbative biasing models.
Thick and thin curves show $b_1$ and $b_2/b_1$ at $z=0$ (solid) and
$z=2$ (dashed).  We use spherical halo model ({\it left}), ellipsoidal
halo model ({\it center}), and peak model ({\it right}).  The biasing
coefficients are plotted against halo mass $M_{\rm halo}$ ({\it left}
and {\it center}), and against peak height $\nu$ ({\it right}).  }
\label{fig:mass_bias}
\end{figure}

Figure \ref{fig:mass_bias} shows biasing coefficients, $b_1$(thick
lines) and $b_2/b_1$(thin lines) at $z=0$ (solid) and $z=2$ (dashed).
We adopt the cold dark matter transfer function of \citet{BBKS1986} as
the initial mass power spectrum (with $\Omega_m=0.3$,
$\Omega_\Lambda=0.7$, and the shape parameter $\Gamma=\Omega_m h=0.21$).
In the three biasing models, $b_1$ and $b_2/b_1$ behave similarly as 
functions of $M_{\rm halo}$ or $\nu$. This implies a presence of a
certain correlation between $b_1$ and $b_2/b_1$.
\begin{figure}[!t]
   \centering \FigureFile(80mm,80mm){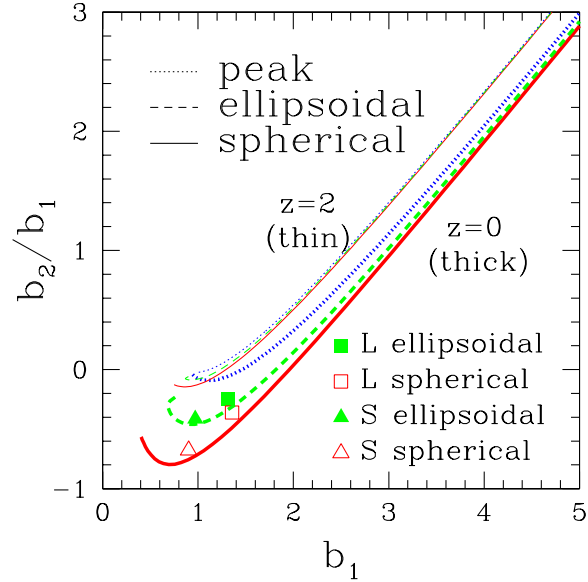}
\caption{Correlation of $b_1$ and $b_2/b_1$ from halo and peak biasing.
Different line types are results of halo biasing of spherical
collapse(solid), ellipsoidal collapse(dashed), and peak biasing (dotted),
evaluated at redshifts 2 (thin) and 0 (thick). Open (filled) symbols
represent the mass-averaged values, $B_1$ and $B_2/B_1$, defined in
equation (\ref{eq:b_ave}), assuming spherical (ellipsoidal) halo
model. The mass ranges correspond to L (square), and S (triangle)
defined in subsection \ref{sec:simulation_SDSS}.  } \label{fig:b1_b2}
\end{figure}
To see this point more clearly, we plot $b_2/b_1$ in terms of $b_1$ for
each model (figure \ref{fig:b1_b2}).  All the three models exhibit very
similar correlations between $b_1$ and $b_2/b_1$. 
The differences among the three models become even smaller at higher 
redshifts; compare $z=0$ (thick lines) and $z=2$ (thin lines).
 
In analyzing simulation data below, the estimate of the biasing
coefficients is made after averaging over a finite mass range.  We
calculate the mass-averaged values of $b_1$ and $b_2$ for halo models at
$z=0$:
\begin{eqnarray}
B_n(m_{\rm min},m_{\rm max}) &=& \frac{\displaystyle\int_{m_{\rm min}}
^{m_{\rm max}}dm\, n_{\rm halo}(m,z=0)b_n(m,z=0)}
{\displaystyle\int_{m_{\rm min}}^{m_{\rm max}} dm\,n_{\rm halo}(m,z=0)}, 
\qquad (n=1,2). 
\label{eq:b_ave}
\end{eqnarray}
The difference between $b_n$ and $B_n$ is illustrated in figure
\ref{fig:b1_b2}, where spherical (open) and ellipsoidal (filled) halo
models [eqs.(\ref{eq:PS}), (\ref{eq:ST})] are assumed. We plot the
results for two mass ranges, corresponding to S (triangle) and L
(square) halo subsamples (see subsection \ref{sec:simulation_SDSS} for
detail).

\begin{figure}[!t]
   \centering \FigureFile(160mm,80mm){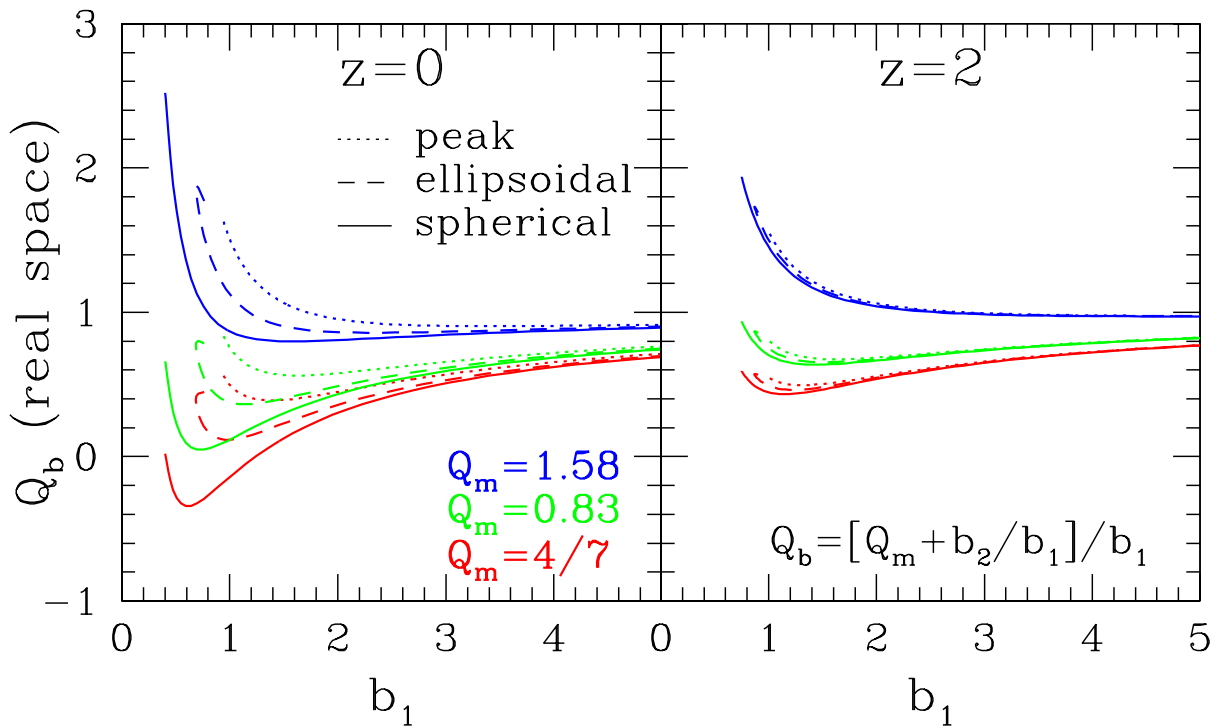}   
\caption{$Q$ for biased fields as a function of $b_1$ in real space.  We
   estimate $Q$ using equation (\ref{eq:Qb}) on the basis of
   perturbative predictions for peaks (dotted), ellipsoidal halos
   (dashed), and spherical halos (solid). We assume three different
   values for $Q_m$: $Q_m=1.58$, $Q_m=0.83$, and $Q_m=4/7$ from top to
   bottom. The first two values are computed from N-body results at
   $k=0.4h{\rm Mpc}^{-1}$, and $k=0.18h{\rm Mpc}^{-1}$ (see section
   \ref{sec:simulation} below), and the last value corresponds to
   perturbation theory [eqs.(\ref{eq:Qtreelevel}) and (\ref{eq:F2})],
   where equilateral triangles ($k_1=k_2=k_3=k$) are assumed.  {\it
   left}: $z=0$, {\it right}: $z=2$.}  \label{fig:b1_Qb}
\end{figure}
\begin{figure}[!t]
   \centering \FigureFile(160mm,80mm){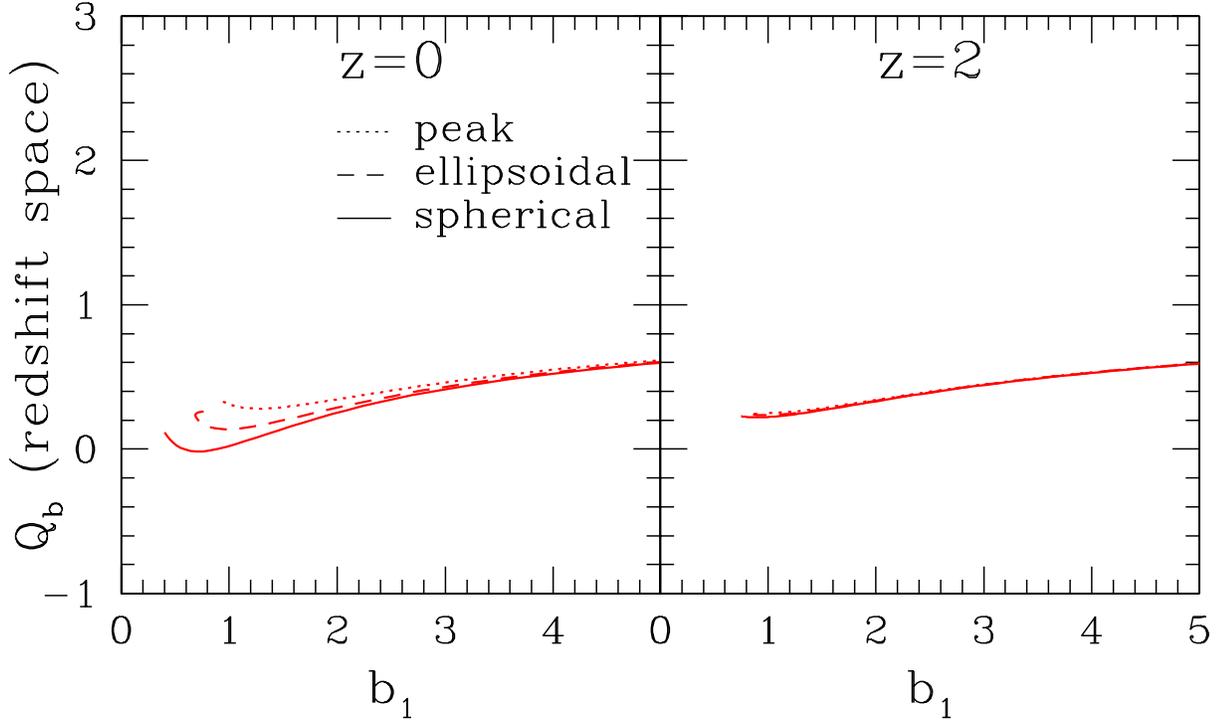}
\caption{Same as figure \ref{fig:b1_Qb}, but for redshift space. We use
   equation (\ref{eq:Qred}) for equilateral triangles instead of
   equation (\ref{eq:Qb}).  In this time, $Q_b$ is calculated directly
   (not in terms of $Q_m$), so we plot only one set corresponding to
   $Q_m=4/7$ in figure \ref{fig:b1_Qb}.  } \label{fig:b1_Qb_red}
\end{figure}

Figure \ref{fig:b1_Qb} plots $Q_b$ for biased fields as a function of
$b_1$ in real space. The dotted, dashed and solid lines represent the
results of tree-level perturbation with respect to biasing, equation
(\ref{eq:Qb}), for peak, ellipsoidal, and spherical halo models.  For
simplicity, we consider equilateral triangles and adopt three different
values for $Q_m$; $Q_m=4/7$ from the leading term of the gravitational
nonlinearity [see eq.(\ref{eq:Qtreelevel}) below], and $Q_m=0.83$ and
$1.58$ from N-body results at $k=0.18h{\rm Mpc}^{-1}$ and $k=0.40h{\rm
Mpc}^{-1}$.

Figure \ref{fig:b1_Qb_red} plots $Q_b$ for biased fields but in redshift
space.  The redshift distortion effect for $Q$ is difficult to model
accurately, and we follow \citet{SCF99} and replace equation
(\ref{eq:Qb}) by
\begin{eqnarray}
Q_{b,s{\rm eq}} &=&
\frac{5(2520+ 4410\gamma + 1890\beta + 2940\gamma\beta +
378\beta^2 + 441\gamma\beta^2 + 9\beta^3 + 1470b_1\beta
+ 882b_1\beta^2- 14b_1\beta^4 ) }{98b_1 (15+10\beta+
3\beta^2 )^2},
\label{eq:Qred}
\\
\beta &\equiv& \frac{1}{b_1}\left[\Omega_m^{4/7}+\frac{\Omega_\Lambda}{70} 
\left(1+\frac{\Omega_m}{2}\right)\right], 
\label{eq:beta}
\end{eqnarray}
for equilateral triangles, with $\gamma\equiv b_2/b_1$. 

These simple halo and peak biasing models imply that the nonlinearity of
biasing weakens the dependence of $Q_{\rm halo}$ and $Q_{\rm peak}$ on
$b_1$, which qualitatively explains the results of \citet{Kayo2004} 
for equilateral triangles. To
be more realistic, however, we consider in the next section nonlinear
gravity, redshift-space distortion and finite survey boundary effect
using N-body simulations and SDSS data.

\section{Comparison with N-body simulations and SDSS galaxies}
\label{sec:simulation}

\subsection{N-body simulations and SDSS galaxies}
\label{sec:simulation_SDSS}

N-body simulations that we use employ $512^3$ dark matter particles in a
cubic box of 300$h^{-1}$Mpc (comoving) on a side with the periodic
boundary condition.  The mass of each dark matter particle is $1.68
\times 10^{10} h^{-1}M_\odot$.  As in section \ref{sec:AM}, we use the
cold dark matter transfer function of \citet{BBKS1986} as the initial
mass power spectrum (with $\Omega_m=0.3$, $\Omega_\Lambda=0.7$, and 
$\Gamma=\Omega_m h=0.21$), and have generated three independent Gaussian
realizations. The initial density field at $z=36$ is evolved up to $z=0$
using the P$^3$M code of Jing \& Suto (1998, 2002) with the gravitational
softening length of $\epsilon \sim 58 h^{-1}$kpc. We assume the amplitude 
of the density fluctuations smoothed over $8h^{-1}{\rm Mpc}$ $\sigma_8 = 0.9$.

\begin{table}[!t]
\caption{Volume-limited subsamples of SDSS galaxies for $0.05<z<0.1$.
}
\begin{center}
\begin{tabular}{cccccccc}
\hline
  subsample & luminosity & color & $N_g$ & $n_g[h^3{\rm Mpc}^{-3}]$ &
  $2\pi n_g^{1/3}[h{\rm Mpc}^{-1}]$ & $b_1$ \\ 
\hline
blue & $-21.3 < {\rm M_r} < -19.8$ & ${\rm g-r} < 0.86$ & 33986 &  0.00359 & 0.962 & 0.95
  \\ 
  red & $-21.3 < {\rm M_r} < -19.8$ & ${\rm g-r} > 0.86$ & 34351 &  0.00363 & 0.966 & 1.33
\\ \hline
\end{tabular}
\end{center}
\label{tab:galaxy_catalog}
\end{table}

In order to test the dependence of $Q_b$ on actual galaxy properties, we
construct volume-limited subsamples of different colors from New York
University Value-Added Galaxy Catalog \citep{Blanton2005} based on the
SDSS Data Release 4 \citep{DR4}. The angular selection function of the
survey is written in terms of spherical polygons \citep{Hamilton2004}.
Details of the SDSS can be found in the following papers:
\citet{York2000} describe an overview, technical articles providing
details include descriptions of the telescope design \citep{Gunn2006},
the photometric camera \citep{Gunn1998}, photometric analysis
\citep{Stoughton2002}, the photometric system and calibration
\citep{Fukugita1996,Hogg2001,Ivezic2004,Smith2002,Tucker2006}, the
photometric pipeline \citep{Lupton2001}, astrometric calibration
\citep{Pier2003}, selection of the galaxy spectroscopic samples
\citep{Eisenstein2001,Strauss2002}, and spectroscopic tiling
\citep{Blanton2003}.

We only consider galaxies with redshifts $0.05\leq z \leq 0.1$ and
magnitudes $-21.3\leq M_{r}\leq -19.8$ (68337 galaxies in total, over
$\sim 4259 {\rm deg}^2$). We divide those galaxies according to their
colors so that each subsample contains roughly the same number: {\it
red} subsample has $g-r>0.86$ (34351 in total), and {\it blue} subsample
has $g-r<0.86$ (33986). In Table \ref{tab:galaxy_catalog}, $N_g$ and
$n_g$ denote the total number of galaxies and the galaxy number density
in each subsample, and the wavenumber corresponding to the mean galaxy
separation is $2\pi n_g^{1/3}$. The linear biasing coefficient,
$b_1$, is evaluated at $k=0.126h{\rm Mpc}^{-1}$ assuming the same set of 
cosmological parameters used in N-body simulations. 
In particular the estimated value of $b_1$ is sensitive to $\sigma_8$. 
The value of $\sigma_8$ is still in controversy between different 
observations. For example WMAP3 gives $\sigma_8 = 0.742\pm0.051$ 
\citep{Spergel2006}, while 2dFGRS gives $0.88{+0.12 \atop -0.08}$ 
\citep{Gaztanaga2005}. Thus in principle our estimated values of 
$b_1$ would increase by about 20\% if we had adopted WMAP3 result.

\begin{table}[!t]
\caption{Subsamples of simulated halos.}
\begin{center}
\begin{tabular}{ccccccc}
\hline
  subsample & $M[h^{-1}M_\odot]$ & $\langle N_h\rangle$ & 
  $\langle n_h\rangle[h^3{\rm Mpc}^{-3}]$ & $2\pi\langle n_h
\rangle^{1/3}[h{\rm Mpc}^{-1}]$ & $\langle b_1\rangle$
\\ \hline
  r-cube S & $1.68\times10^{12}$ --- $1.18\times10^{13}$ & 
58692 & 0.00217 & 0.814 & 0.83$\pm$0.01 
  \\ 
  r-cube L & $>1.18\times10^{13}$ & 10881 &  0.000403 & 
0.461 & 1.26$\pm$0.02 
  \\ 
  r-cube LL & $>6.72\times10^{13}$ & 1484 &  0.0000550 & 0.239 & 1.8$\pm$0.1 
  \\
  s-cube S & $1.68\times10^{12}$ --- $1.18\times10^{13}$ & 58692 & 
0.00217 & 0.814 & 0.88$\pm$0.01 
  \\ 
  s-cube L & $>1.18\times10^{13}$ & 10881 &  0.000403 & 0.461 & 1.29$\pm$0.01 
  \\ 
  s-cube LL & $>6.72\times10^{13}$ & 1484 &  0.0000550 & 0.239 & 1.8$\pm$0.1 
  \\
  wedge S & $1.68\times10^{12}$ --- $1.18\times10^{13}$ & 21735 & 
0.00230 & 0.829 & 0.84$\pm$0.01
  \\ 
  wedge L & $>1.18\times10^{13}$  & 3979 &  0.000421 & 0.471 & 1.24$\pm$0.09 
  \\
  wedge LL & $>6.72\times10^{13}$ & 551 &  0.0000582 & 0.243 & 1.8$\pm$0.2 
\\ \hline
\end{tabular}
\end{center}
\label{tab:halo_catalog}
\end{table}

In order to compare with analytical predictions for halo biasing and
also with the SDSS subsamples, we identify dark matter halos using a
friends-of-friends algorithm. Specifically we use the public code
``FOF'' \footnote{available at the website
http://www-hpcc.astro.washington.edu/}, and adopt the linking length
0.164 in units of the mean particle separation. We construct nine 
halo subsamples from three realizations (Table \ref{tab:halo_catalog}). 
The indices, S, L, and LL, represent the three different halo mass ranges 
corresponding to $1.68\times10^{12}h^{-1}M_\odot < M <
1.18\times10^{13}h^{-1}M_\odot$, $1.18\times10^{13}h^{-1}M_\odot < M$, 
and $6.72\times10^{13}h^{-1}M_\odot < M$. The mass ranges of the first 
two, S and L, are determined so that they have approximately the same 
values of $b_1$ for the SDSS blue and red galaxies, respectively. 
The last one, LL, is constructed so as to check the $b_2$-$b_1$ correlation 
around $b_1\sim2$, and discussed in figure \ref{fig:b2b1_sim_ana} below. 
We further divide the three different mass-selected samples into 
three subsamples: r-cube (s-cube) measures positions of halos in 
real (redshift) space using the original simulation cube, 
$(300h^{-1}{\rm Mpc})^3$.  In the latter,
redshift-space distortion effect is taken into account by using
$z$-component of the center-of-mass velocity of each halo.  Wedge
subsamples are constructed so as to have the same survey geometry as the
SDSS sample. They measure the positions of halos in redshift space,
using the line-of-sight component of the center-of-mass velocity unlike
s-cube.  In Table \ref{tab:halo_catalog}, $\langle N_h\rangle$ and
$\langle n_h\rangle$ denote the total number of halos and the halo
number density in each subsample (averaged over three realizations). The
wavenumber corresponding to the mean halo separation is $2\pi\langle
n_h\rangle^{1/3}$, and the linear biasing coefficient, $\langle
b_1\rangle$ is evaluated at $k=0.126h{\rm Mpc}^{-1}$.

\subsection{Power spectrum and $b_1$}

\begin{figure}[!t]
   \centering \FigureFile(120mm,120mm){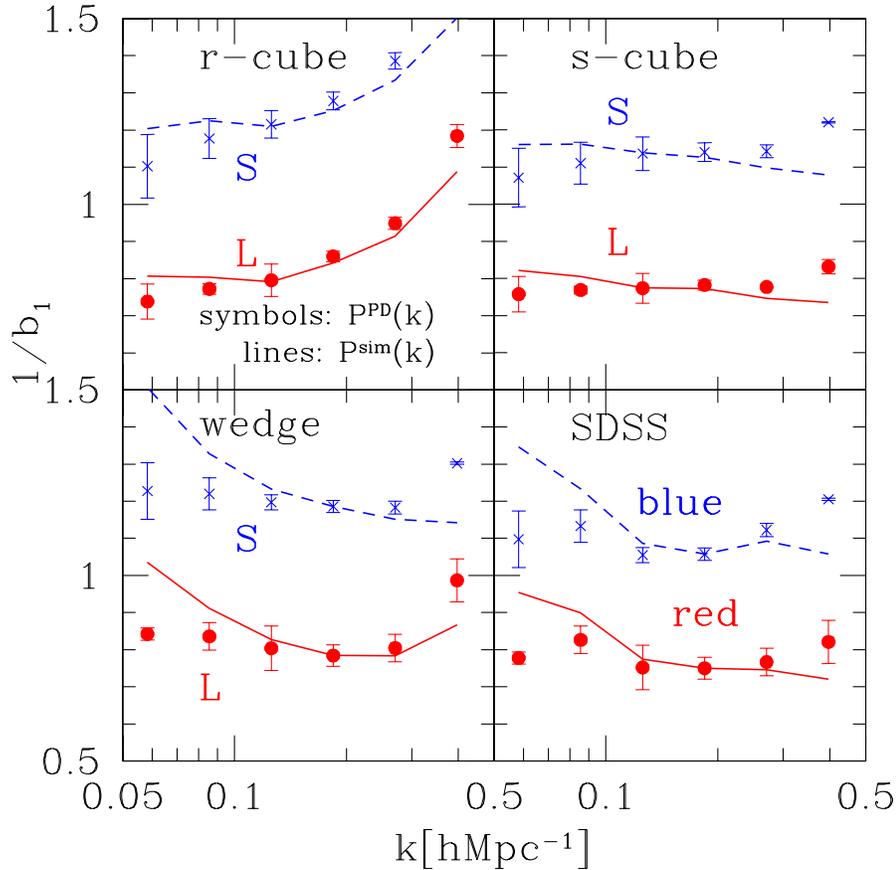} 
\caption{Inverse of the linear biasing parameters of simulated halos
(Table \ref{tab:halo_catalog}) and SDSS galaxies (Table
\ref{tab:galaxy_catalog}). Crosses and filled circles for simulated
halos (SDSS galaxies) correspond to S and L samples (blue and red),
respectively. Dashed and solid lines indicate results based on 
the direct estimation of $P(k)$, while symbols in all panels 
use $P^{\rm PD,r}(k)$ or $P^{\rm PD,s}(k)$, equation (\ref{eq:pks0}). 
The quoted error bars for simulated halos are computed from 
three different realizations. We simply use the error bars
for wedge subsamples just for reference in the case of SDSS galaxies. }
\label{fig:binv}
\end{figure}
First we estimate the linear biasing parameter according to
equation (\ref{eq:pb}): 
\begin{equation}
b_{1,i}(k) = \sqrt{\frac{P_{i}(k)}{P(k)}},
\label{eq:binv}
\end{equation}
where the subscript $i$ denotes the different subsamples of simulated halos 
and SDSS galaxies. 

In real-space (r-cube in Table \ref{tab:halo_catalog}), we first
construct density field of halos using the cloud-in-cell pixelization,
Fourier transform it, angularly average $P_{i}({\bf k})$ over the
direction of ${\bf k}$, and finally remove the shot noise contribution,
$1/n_h$ to obtain $P_{i}(k)$.  We choose logarithmically equal bins for
$k$, $\Delta(\log_{10} k) = 1/6$. The mean values $\langle n_h\rangle$ of the
number density of halos in each realization ($n_h$) are listed in Table
\ref{tab:halo_catalog}. The wavenumbers corresponding to the mean halo
separation $2\pi/\langle n_h\rangle^{1/3}$ roughly provide the limit of
the reliability of estimating $P_{i}(k)$.  The dark matter power
spectrum $P(k)$ is estimated in two different methods.  One is a direct
estimate using the dark matter particles, $P^{\rm sim,r}(k)$. The other
employs an analytical prescription by \citet{Peacock1996}, $P^{\rm
PD,r}(k)$. Upper-left panel of figure \ref{fig:binv} shows these
results. Symbols and curves represent the results using $P^{\rm
PD,r}(k)$ and $P^{\rm sim,r}(k)$. Their agreement ensures the validity
of the Peacock-Dodds prescription. 

The wavenumber dependence of $1/b_1$ is weak for $k<0.2h{\rm Mpc}^{-1}$.
The increase of $1/b_1$ for $k>0.2h{\rm Mpc}^{-1}$ is consistent with 
that expected from the halo finite volume exclusion effect (e.g., 
\cite{Taruya2001}). 
When the wavenumber becomes comparable or larger than the mean separation 
of halos ($\sim 0.5h{\rm Mpc}^{-1}$; c.f. Table \ref{tab:halo_catalog}), 
$P_{i}(k)$ does not represent the intrinsic clustering signal properly. 
Therefore the increase of $1/b_1$ against $k$ may not be real for 
$k > 0.5h$Mpc$^{-1}$. In this paper we are interested in linear regimes 
where the $k$ dependence of $b_1$ is negligible. Thus we do not consider 
the region $k>0.2h$Mpc$^{-1}$, where the scale independent expansion 
like equation (\ref{eq:db-d}) breaks down.

In redshift space (s-cube in Table \ref{tab:halo_catalog}), we compute
$P_{i}(k)$ using the center-of-mass peculiar velocity of each halo
adopting the distant observer approximation. The dark matter power
spectrum in {\it redshift space} is computed with two different
methods.  One is a direct estimate using the dark matter particles,
$P^{\rm sim,s}(k)$. The other, $P^{\rm PD,s}(k)$, is an analytical
prediction combining $P^{\rm PD,r}(k)$ and redshift distortion effects
empirically. For the latter we consider the Kaiser effect
\citep{Kaiser1987} and the finger-of-God effect assuming exponential
peculiar velocity distribution \citep{Cole1994,Cole1995,Kang2002}. 
In this case the redshift space power spectrum is given as
\begin{eqnarray}
\label{eq:pks0}
  P^{\rm PD,s}(k) &=& 
\left[A(\kappa_m)+{2\over3}\beta_m B(\kappa_m)+{1\over 5}\beta_m^2 
C(\kappa_m)\right]  P^{\rm PD,r}(k), \\
  A(\kappa_m) &=& {1\over\kappa_m}{\arctan}(\kappa_m),
\\
  B(\kappa_m) &=& {3\over\kappa_m^2}\biggl(1-A(\kappa_m)\biggr),
\\
  C(\kappa_m) &=& {5\over3\kappa_m^2}\biggl(1-B(\kappa_m)\biggr),
\end{eqnarray}
where $\beta_m$ denotes the value of equation (\ref{eq:beta}) for 
dark matter particles, $\kappa_m=k\sigma_p/(\sqrt{2}H_0)$, and the 
rms pairwise velocity dispersion of dark matter particles, 
$\sigma_p$ ($= \sqrt{2}\sigma_v \approx 520{\rm km/s}$, 
$\sigma_v$ is the one-dimensional velocity dispersion), is computed 
directly from N-body simulations. Upper-right panel in figure 
\ref{fig:binv} displays the results for s-cube. Again $b_1$ is
fairly scale independent in redshift space.  Although the finger-of-God
effect partially compensates the wavenumber dependence in r-cube for
$k>0.2h{\rm Mpc}^{-1}$, $b_1(k)$ there is not
reliable. The results of r-cube and s-cube imply that $b_1(k)$ is not
sensitive to the redshift distortion, i.e., $b_1^s\approx b_1^r$ for
$k<0.2h{\rm Mpc}^{-1}$. In fact, the agreement between $P^{\rm sim}(k)$
and $P^{\rm PD}(k)$ suggests that the above feature can be explained
using equation (\ref{eq:pks0}):
\begin{eqnarray}
\frac{b_{1,i}^s}{b_{1,i}^r} &=& \sqrt{\frac{A(\kappa_i)+\frac23\beta_iB(\kappa_i)
+\frac15\beta_i^2C(\kappa_i)}{A(\kappa_m)+\frac23\beta_mB(\kappa_m)
+\frac15\beta_m^2C(\kappa_m)}},
\label{eq:b1rs}
\end{eqnarray}
where subscripts $i$ and $m$ refer to different subsamples and dark
matter particles, respectively, and $\beta_i$ indicates the value of equation 
(\ref{eq:beta}) for each subsample, and $\kappa_i=k\sigma_{p,i}/(\sqrt{2}H_0)$.
The rms pairwise velocity dispersion of halos, $\sigma_{p,i}$, is 
calculated from the simulated halo subsamples through 
$\sigma_p=\sqrt{2}\sigma_v$; 412 km/s for S and 397 km/s for L. 
Equation (\ref{eq:b1rs}) yields 1.04 (0.97) at $k=0.05h{\rm Mpc}^{-1}$
and 1.08 (1.01) at $k=0.2h{\rm Mpc}^{-1}$ for the S (L) subsample.
These values explain the behavior in figure \ref{fig:binv} and suggest
that $b_1^s$ and $b_1^r$ in linear regime agree within 10\% level.

For wedge subsamples, we determine the positions of halos in redshift
space properly using the line-of-sight velocity component unlike s-cube
where we employ the distant observer approximation.  In estimating
$P_i(k)$, we follow \citet{FKP1994}, \citet{Matarrese1997} and
\citet{Verde2002}, and define the field:
\begin{eqnarray}
F_i({\bf r}) \equiv \lambda w({\bf r})[n_i({\bf r})-\alpha n_{\rm r}({\bf r})],
\end{eqnarray}
where $w({\bf r})$ is the weight, $\lambda$ is a constant to be
determined, $n_{\rm r}({\bf r})$ is the number density of random
particles, $n_i({\bf r})$ is the number density of each subsamples, and
$\alpha$ is the ratio of particle numbers of actual and random catalogs.
In this paper, $w({\bf r})$ is unity inside the survey volume, and zero
otherwise. If we set $\lambda=I_{22}^{-1/2}$, where
\begin{eqnarray}
I_{jk} = \int{\rm d}^3r\,w^j({\bf r})\bar{n}_i^k({\bf r})
\end{eqnarray}
with the mean number density $\bar{n}_i$ for each subsample
\citep{Matarrese1997}, then the power spectrum is
\begin{eqnarray}
\langle|F_i({\bf k})|^2\rangle = P_i(k)+\frac{I_{21}}{I_{22}}(1+\alpha)
\label{eq:P_estimate}
\end{eqnarray}
\citep{Verde2002}.  
The dark matter power spectrum for the wedge
subsamples is calculated by the two methods again: the 
first is an analytical calculation using equation (\ref{eq:pks0}) 
and the second is a direct estimation based on equation 
(\ref{eq:P_estimate}). The results are plotted in the lower-left 
panel of figure \ref{fig:binv}. 
Symbols correspond to the estimate using equation (\ref{eq:binv}) 
with $P(k)$ evaluated from the Peacock-Dodds prescription, i.e., 
$P^{\rm PD}(k)$. Solid and dashed lines use the direct estimate 
for $P(k)$, instead, i.e., $P^{\rm sim}(k)$. 
The symbols and the lines deviate for $k < 0.1h$Mpc$^{-1}$ 
where the effect of complicated survey volume shape cannot be ignored.
For $k>0.1h{\rm Mpc}^{-1}$ the estimate of $b_1$ in wedge 
configuration well reproduces that in s-cube, upper-right panel 
of figure \ref{fig:binv}.

Finally $b_1(k)$ for SDSS galaxies is computed similarly as wedge
subsamples. Again we assume the same 
cosmological parameters used in the N-body simulations. 
As in the wedge case, we use $P(k) = P^{\rm PD}(k)$ and $P^{\rm sim}(k)$ 
for symbols and lines, respectively.
The lower-right panel of figure \ref{fig:binv} 
is the result.  The red (blue) sample is almost the same as L (S) in the 
lower-left panel. Lower-left panel of figure \ref{fig:binv} is indeed in good
agreement with the upper panels for $k<0.2h{\rm Mpc}^{-1}$, which we may
interpret as an indication that the SDSS results in redshift space are
directly related to their real-space property.
We find very similar deviations between lines and symbols at 
$k < 0.1$, which ensures our interpretation due to the effect 
of complicated survey boundary shape.

\subsection{Bispectrum and $Q$}
\begin{figure}[!t]
	\centering \FigureFile(120mm,120mm){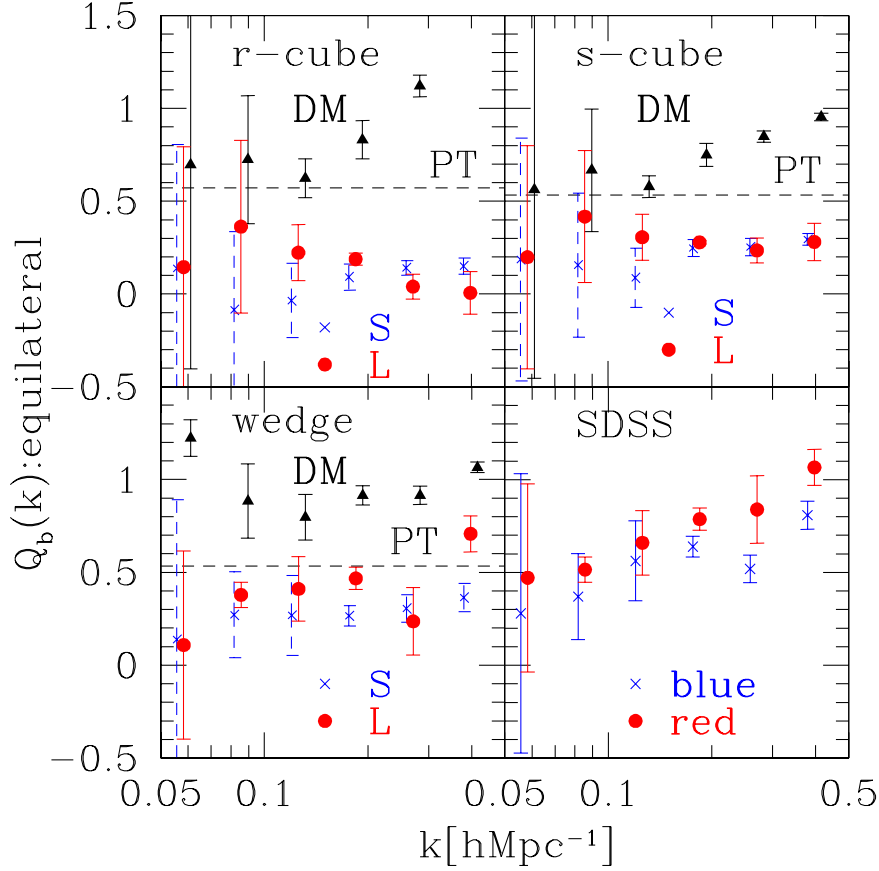}
\caption{$Q_{halo}$ for simulated halos (Table \ref{tab:halo_catalog})
and $Q_{galaxy}$ for SDSS galaxies (Table
\ref{tab:galaxy_catalog}). Crosses and filled circles for simulated
halos (SDSS galaxies) correspond to S and L samples (blue and red),
respectively. For comparison, we also plot the value of $Q_m$
calculated from N-body simulations (filled triangles) and from
perturbation theory (PT, dashed lines) in r-cube, s-cube and wedge. 
The quoted error bars for simulated halos and dark matters are 
computed from three different realizations. We simply use the error 
bars for wedge subsamples just for reference in the case of SDSS 
galaxies.}
\label{fig:Q}
\end{figure}
Finally we are in a position to consider the three-point statistics. 
We calculate the bispectra $B_i({k_1},{k_2},{k_3})$ using the 
same methods as $P_i(k)$ for r-cube and s-cube. For wedge and 
SDSS subsamples we use the formula 
\begin{equation}
\langle F_i({\bf k_1})F_i({\bf k_2})F_i({\bf k_3})\rangle=\frac{I_{33}}
{I_{22}^{3/2}}\left\{B_i({\bf k_1},{\bf k_2},{\bf k_3})+\frac{I_{32}}
{I_{33}}\left[P_i({\bf k_1})+P_i({\bf k_2})+P_i({\bf k_3})\right]+
(1-\alpha^2)\frac{I_{31}}{I_{33}}\right\}
\end{equation}
\citep{Verde2002}. 
Figure \ref{fig:Q} plots $Q_i(k)$, the reduced amplitude of bispectrum for 
equilateral triangles 
\begin{eqnarray}
Q_i(k) \equiv \frac{B_i(k,k,k)}{3P^2_i(k)},
\label{eq:Qi}
\end{eqnarray}
which should be compared with figure \ref{fig:binv}. 
The difference of $b_1$ between the two different subsamples in 
each panel of figure \ref{fig:binv} does 
not show up in figure \ref{fig:Q}. This is fully consistent with the
finding of \citet{Kayo2004} for three-point correlation
functions. Furthermore comparison among simulated halo subsamples
indicates that this feature is not a simple outcome of the redshift 
distortion effect, but reflects the intrinsic correlation between $b_1$ 
and $b_2/b_1$ in real space as we discussed in section \ref{sec:AM}. 

In order to proceed further, we attempt to estimate $b_2(k)/b_1(k)$ 
combining $Q_i(k)$ and $Q_m(k)$. We directly compute $Q_m(k)$ from 
simulation particles for r-cube, s-cube and wedge (filled triangles 
in fig.\ref{fig:Q}). For comparison, tree-level perturbation theory 
predicts that
\begin{equation}
Q(k_1,k_2,k_3) = 2\frac{F_2({\bf k_1},{\bf k_2})P(k_1)P(k_2)
+F_2({\bf k_2},{\bf k_3})P(k_2)P(k_3)+
F_2({\bf k_3},{\bf k_1})P(k_3)P(k_1)}{P(k_1)P(k_2)+P(k_2)P(k_3)+P(k_3)P(k_1)},
\label{eq:Qtreelevel}
\end{equation}
where we follow \citet{JB94} and define $F_2({\bf k_1},{\bf k_2})$ as 
\begin{equation}
\label{eq:F2}
F_2({\bf k_1},{\bf k_2}) = \frac{5}{7}
+\frac{1}{2}\frac{{\bf k_1}\cdot{\bf k_2}}{k_1k_2}
\left(\frac{k_1}{k_2}+\frac{k_2}{k_1}\right)
+\frac{2}{7}\frac{({\bf k_1}\cdot{\bf k_2})^2}{k_1^2k_2^2}.
\end{equation}
Strictly speaking, this expression of $F_2({\bf k_1},{\bf k_2})$ is
valid only for the Einstein-de Sitter universe, but its dependence on
cosmology is proved to be very weak \citep{matsubara95,scoccimarro98}.
For equilateral triangles, equation (\ref{eq:Qtreelevel}) reduces to a
constant value, $4/7$, which is plotted as a dashed line (PT) in the
upper-left panel of figure \ref{fig:Q} for real space. In redshift
space, one obtains $Q=0.533$ by setting $\gamma=0$ and $\beta=\beta_m$
in the perturbation expression, equation (\ref{eq:Qred}), which is also
plotted as a dashed line (PT) in the upper-right and lower-left panels. 
For $k<0.2h{\rm Mpc}^{-1}$, the direct estimates (filled triangles; DM)
agree with the perturbation values.  In directly evaluating $Q$ from
N-body simulations, we use equation (\ref{eq:Qm}) for $Q_m$, and a
similar expression for $Q_i$.

\begin{figure}[!t]
	\centering \FigureFile(120mm,120mm){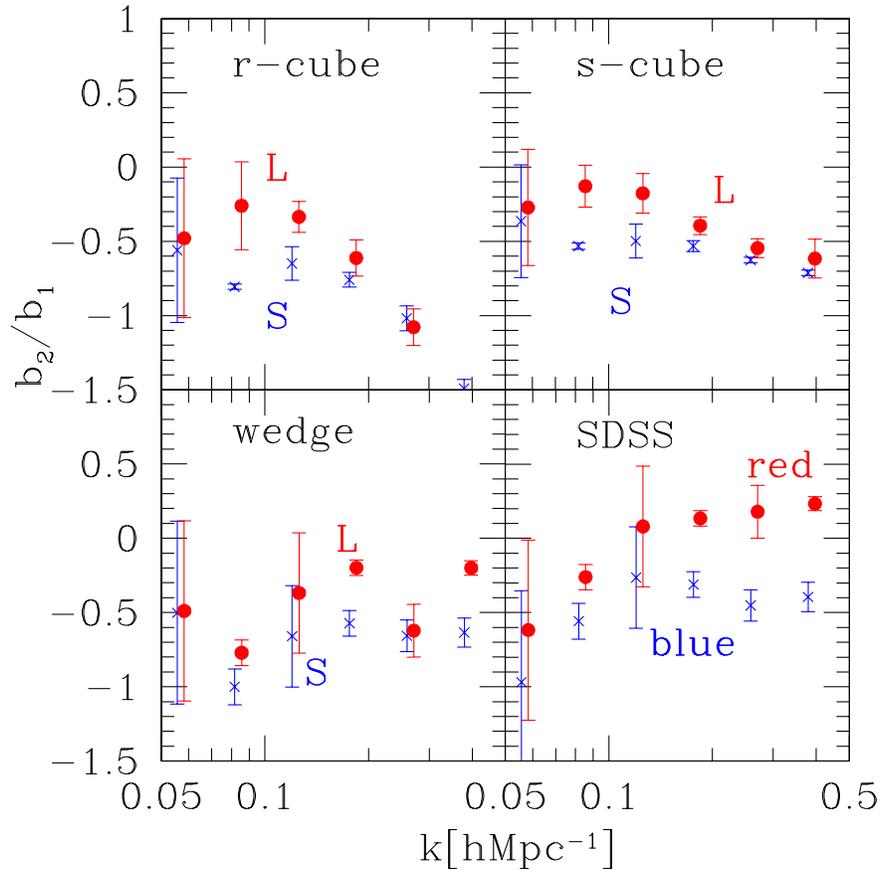}
\caption{$b_2/b_1$ for simulated halos (Table \ref{tab:halo_catalog})
and SDSS galaxies (Table \ref{tab:galaxy_catalog}). 
Crosses and filled circles for simulated 
halos (SDSS galaxies) correspond to S and L samples (blue and red),
respectively. The quoted error bars for simulated halos are computed 
from three different realizations. We simply use the error bars for wedge
subsamples just for reference in the case of SDSS galaxies.}
\label{fig:b2b1}
\end{figure}
Figure \ref{fig:b2b1} plots $b_2/b_1$ computed from 
\begin{eqnarray}
(b_2/b_1)_i(k) \equiv b_{1,i}(k)
Q_i(k)-Q_m(k).
\label{eq:b2b1_Q}
\end{eqnarray}
For SDSS subsamples we use the value evaluated from wedge for
$Q_m(k)$.  Strictly speaking, equation (\ref{eq:b2b1_Q}) ignores
redshift-space distortion effects. In reality, however, we made sure
that almost the same values are derived even when we use equation
(\ref{eq:Qred}) instead. Therefore we use equation (\ref{eq:b2b1_Q})
both in real- and redshift-space. Comparison between figure
\ref{fig:binv} and figure \ref{fig:b2b1} indicates that subsamples with
larger $b_1$ tend to have larger $b_2/b_1$.  This is consistent
qualitatively with analytical results shown in figure \ref{fig:b1_b2}.

\begin{figure}[!t]
  \centering \FigureFile(80mm,80mm){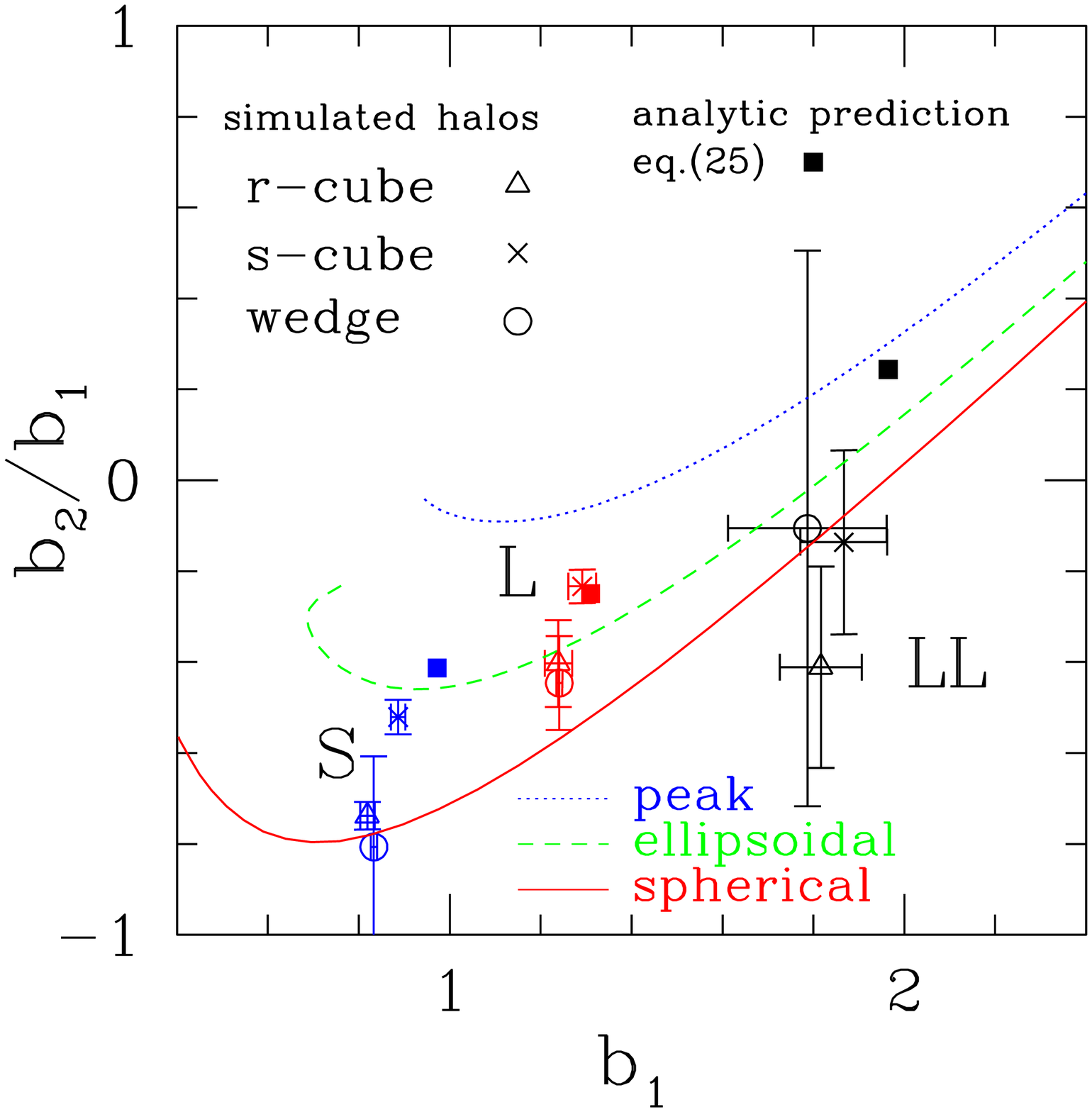}
  \centering \FigureFile(80mm,80mm){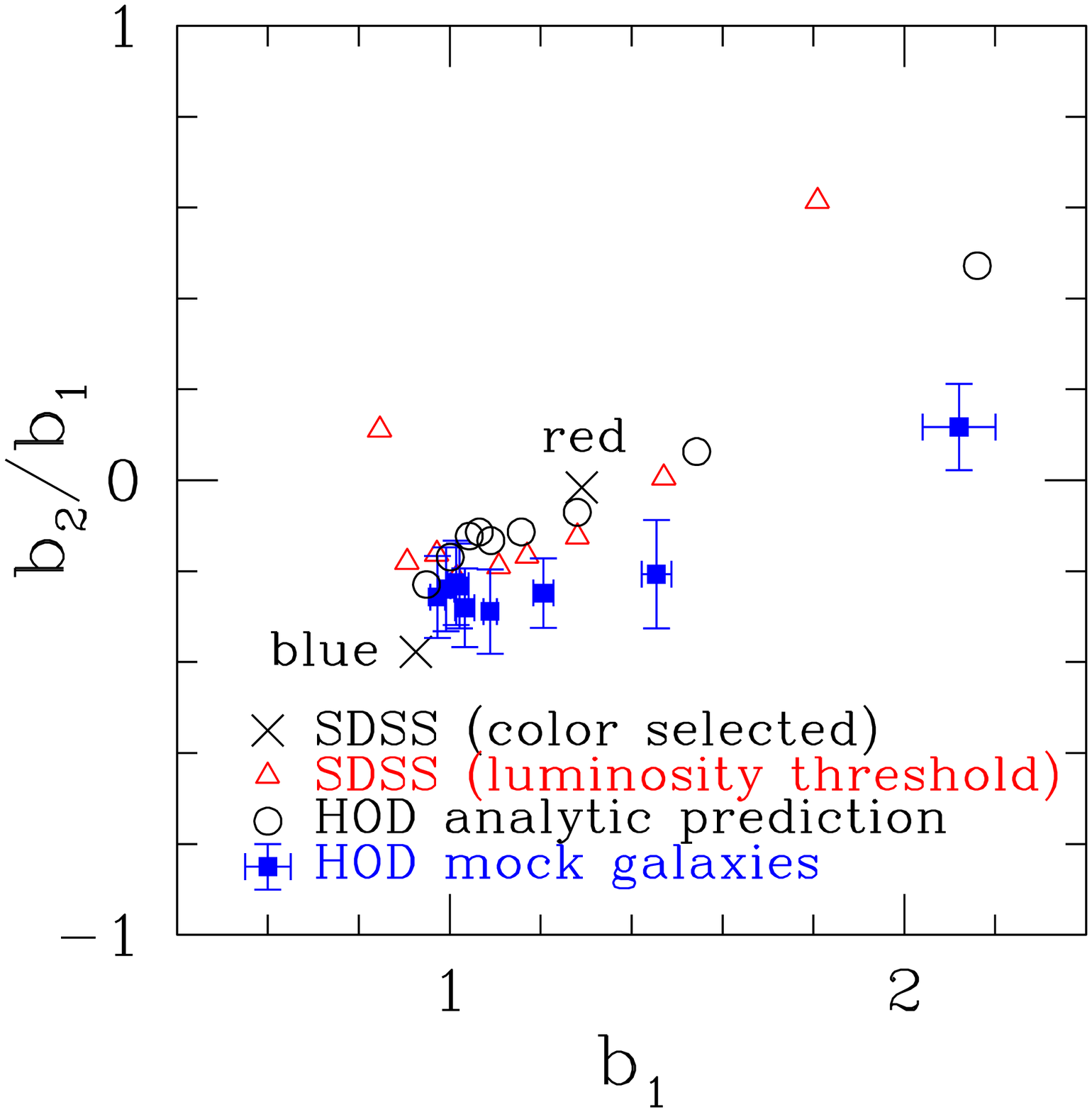}
  \caption{Correlation between $b_1$ and $b_2/b_1$.
    {\it left}: analytic models and simulated halos 
    (Table \ref{tab:halo_catalog}). Lines are analytical 
    predictions for spherical (solid), ellipsoidal (dashed) halo models, 
    and peak model (dotted) as in figure \ref{fig:b1_b2} at $z=0$. 
    Squares are the mass averaged values corresponding to S and L subsamples 
    for the ellipsoidal halo model. Other symbols are for halo 
    subsamples. The quoted error bars are computed from three different
    realizations. 
    {\it right}: SDSS color selected samples 
    (Table \ref{tab:galaxy_catalog}; crosses), 
    SDSS luminosity threshold samples 
    (Table \ref{tab:luminosity_threshold}; triangles), 
    HOD mock galaxies (Table \ref{tab:luminosity_threshold}; squares), 
    and HOD analytic prediction (circles).
    Note that we do not distinguish $b_n$ and $B_n$ here.}
  \label{fig:b2b1_sim_ana}
\end{figure}
To compare more quantitatively simulation results with analytic
models, we average $b_{1,i}(k)$ and $(b_2/b_1)_i(k)$ over the range of
$0.08h{\rm Mpc}^{-1}<k<0.2h{\rm Mpc}^{-1}$. 
 The results are plotted in 
the left panel of figure \ref{fig:b2b1_sim_ana}.  The three curves show 
the analytic models plotted in figure \ref{fig:b1_b2} at $z=0$, while 
squares correspond to the mass-averaged values for ellipsoidal 
halo model again in figure \ref{fig:b1_b2}. Triangles, crosses, and 
circles, represent r-cube, s-cube, and wedge subsamples, respectively. 
If the analytic ellipsoidal halo model is exact, triangles and squares 
should agree within the error bars (we estimate from variance for three 
different realizations).  The small differences between them may be 
ascribed to i) the inaccuracy of the higher order biasing coefficients 
in the halo model. The parameters ($p$ and $q$) in equation (\ref{eq:ST}) 
are empirically determined so as to reproduce the mass function but they 
do not guarantee the reliability for $b_2$. ii) Weak scale dependence which 
is seen in figures \ref{fig:binv} and \ref{fig:b2b1}.  This is not expected 
in the analytic model, implying its practical limitation.  iii) Inaccuracy 
of the estimators of $b_1$ and $b_2/b_1$, and iv) stochasticity of biasing
\citep{Dekel1999,Taruya2000,Taruya2001,Yoshikawa2001}. The present halo
model assumes deterministic biasing, which may affect the mean values of
biasing coefficients, especially in the higher order.
Given those realistic issues, we would interpret that triangles and
squares are in reasonable agreement.

\begin{table}[!t]
\caption{HOD parameters corresponding to the SDSS luminosity threshold 
galaxies}
\begin{center}
\begin{tabular}{cccccccc}
\hline
  $M_r^{\rm max}$ & $z_{\rm min}$ & $z_{\rm max}$ & 
$N_{\rm gal}^{\rm SDSS}$ & $\log_{10}M_{\rm min}$ & 
$\log_{10}M_{1}$ & $\alpha$ & $\langle N_{\rm gal}^{\rm mock}\rangle$ 
\\ \hline
  -22.0 .......... & 0.02 & 0.22 & 7704 & 13.91 & 14.92 & 1.43 & 1417
  \\ 
  -21.5 .......... & 0.02 & 0.19 & 24711 & 13.27 & 14.60 & 1.94 & 7235
  \\ 
  -21.0 .......... & 0.02 & 0.15 & 41969 & 12.72 & 14.09 & 1.39 & 28151
  \\ 
  -20.5 .......... & 0.02 & 0.13 & 69217 & 12.30 & 13.67 & 1.21 & 75292
  \\ 
  -20.0 .......... & 0.02 & 0.10 & 63770 & 12.01 & 13.42 & 1.16 & 141424
  \\ 
  -19.5 .......... & 0.02 & 0.08 & 56384 & 11.76 & 13.15 & 1.13 & 246421
  \\ 
  -19.0 .......... & 0.02 & 0.06 & 30532 & 11.59 & 12.94 & 1.08 & 365719
  \\ 
  -18.5 .......... & 0.02 & 0.05 & 24636 & 11.44 & 12.77 & 1.01 & 514269
  \\ 
  -18.0 .......... & 0.02 & 0.04 & 16123 & 11.27 & 12.57 & 0.92 & 745340
\\ \hline
\end{tabular}
\end{center}
\label{tab:luminosity_threshold}
\end{table}
The right panel of figure \ref{fig:b2b1_sim_ana} plots $b_2/b_1$ 
against $b_1$ for SDSS galaxy subsamples, which should be compared 
with the left panel for simulated halos. In addition to the color 
selected subsamples in Table \ref{tab:galaxy_catalog}, we construct 
nine luminosity threshold subsamples 
(Table \ref{tab:luminosity_threshold}) following \citet{Zehavi2005}. 
We construct realistic galaxy mock samples in order to see if the 
observational data can be understood from simple theoretical modeling. 
We employ halo occupation distribution 
approach (HOD) which assigns galaxies within simulated halos. 
According to the simplest version of HOD, mean number of galaxies 
within halos of mass $M$ is set as 
\begin{eqnarray}
  \langle N(M)\rangle = \left\{
  \begin{array}{ll}\displaystyle
    1+\left(\frac{M}{M_1}\right)^\alpha&\quad M>M_{\rm min},\\
    0 &\quad {\rm otherwise}.
  \end{array}
  \right.
\label{eq:HOD}
\end{eqnarray}
In the above expression, the first term represents a central galaxy 
while the second corresponds to satellite galaxies. 
The three parameters, $M_{\rm min}$, $M_1$, and $\alpha$, are 
determined to reproduce the observed sample of galaxies. In 
practice, we adopt the values (Table \ref{tab:luminosity_threshold}) 
fitted by \citet{Zehavi2005} for the SDSS luminosity galaxies 
(see their Table 2). 
Their fit assumes the density profile 
proposed by \citet{Navarro1996} for galaxy distribution in each 
halo, and satellite galaxies within each halo are assigned following 
the Poisson distribution of the mean value of equation (\ref{eq:HOD}). 
We construct mock galaxy samples from our simulated halo catalogs 
using the routine of 
\citet{Skibba2006} which simultaneously takes care of the 
redshift-space distortion effect (see also \cite{Kang2005}). 
In Table \ref{tab:luminosity_threshold}, $N_{\rm gal}^{\rm SDSS}$ and 
$\langle N_{\rm gal}^{\rm mock}\rangle$ denote the numbers of SDSS 
galaxy samples and the mean numbers of mock galaxy samples. 
Since the galaxy samples of \citet{Zehavi2005} are 
based on different galaxy catalog (SDSS DR2; \cite{Abazajian2004}) from 
this work, the galaxy numbers of our luminosity threshold samples are 
about twice larger than theirs. We simply distribute mock galaxies 
in the simulation cubes and do not consider the survey boundary 
effect, so the numbers of mock galaxy samples do not correspond to 
those of SDSS galaxy samples.

According to the current HOD approach, we evaluate the biasing 
coefficients in two ways. One is an analytical estimate (HOD analytic 
prediction) which is based on equation (\ref{eq:b_ave}): 
\begin{eqnarray}
B_n &=& \frac{\displaystyle\int_{M_{\rm min}}^{M_{\rm max}} dM\, 
n_{\rm halo}(M,z=0)\langle N(M)
\rangle b_n(M,z=0)}{\displaystyle\int_{\rm M_{\rm min}}^{\rm M_{\rm max}} 
dM\,n_{\rm halo}(M,z=0)
\langle N(M)\rangle}, 
\qquad (n=1,2), 
\label{eq:b_ave_hod}
\end{eqnarray}
where we use the Sheth-Tormen mass function, equation (\ref{eq:ST}), 
for $n_{\rm halo}(M,z=0)$, and equation (\ref{eq:halo_bias}) 
for $b_n(M,z=0)$, and $M_{\rm max}$ is set as the maximum 
mass of our simulated halos. 
The other is the direct evaluation of the biasing 
coefficients from the HOD mock galaxy samples using equations 
(\ref{eq:binv}) and (\ref{eq:b2b1_Q}). 
The right panel of figure \ref{fig:b2b1_sim_ana} plot these results for 
SDSS color selected samples (crosses), for SDSS luminosity threshold 
samples (triangles), for HOD analytic prediction (circles) and finally 
for HOD mock galaxies (squares with error bars). 
We note a few interesting features in the plot. 
i) The similarity between the color selected and luminosity 
threshold samples. While they are based on different selection 
criteria, the resulting $b_2/b_1$ -- $b_1$ correlation seems to be 
roughly the same. 
ii) The HOD analytic prediction reproduces the values of $b_2/b_1$ 
and $b_1$ for each luminosity threshold galaxy samples. 
This is surprising since the fitting procedure is designed 
to reproduce their two-point correlation functions alone. 
iii) The discrepancy between the HOD mock and analytical results may 
come from the difference between Sheth-Tormen model and our halo samples 
in $n_{\rm halo}$ and/or $b_2$. 
We made sure that $n_{\rm halo}$ of our halo samples is in 
good agreement with equation (\ref{eq:ST}). On the other hand as the left 
panel of figure \ref{fig:b2b1_sim_ana} indicates, $b_2$ for our simulated 
halos is systematically lower than the analytic prediction, equation 
(\ref{eq:b_ave}). 
Thus the agreement between SDSS data and HOD models 
needs to be interpreted with caution. We also compute the values 
of $b_2/b_1$ and $b_1$ of HOD mock galaxy samples in real space. 
We, however, do not plot the results, since they are almost the same 
as those in redshift space. 

Even with the above subtlety we find a fairly generic correlation 
between $b_2/b_1$ and $b_1$, which is the most important result of 
the current study. We note here that \citet{Gaztanaga2005} performed 
a related analysis using 2dF galaxy data. They found that 
$b_1 = 0.93{+0.10 \atop -0.08}$ and $b_2/b_1 = -0.34{+0.11 \atop -0.08}$.
These values are in good agreement with the right panel of figure 
\ref{fig:b2b1_sim_ana}. Incidentally they speculated a crude correlation 
of $b_2/b_1 \sim b_1-1.2$, which is indeed consistent with our finding 
on the basis of systematic results from analytic, and numerical 
simulations and SDSS galaxy data.

\section{Summary and Discussion}
\label{sec:SD}
We have found a fairly generic correlation between linear and 
quadratic biasing coefficients, $b_1$ and $b_2/b_1$, using a 
variety of different methodologies; perturbative expansions of 
peak and halo biasing models, N-body simulations of halos, 
SDSS galaxy data analysis, and the corresponding halo 
occupation distribution predictions (analytic and mock). 
The presence of such correlations was suggested earlier by 
the previous finding that the normalized three-point correlation 
functions of SDSS galaxies in redshift space follow the 
hierarchical relation approximately, $Q=0.5\sim1.0$, despite 
the robust morphological, color and luminosity dependences 
of the corresponding two-point correlation functions
\citep{Kayo2004}. We have derived the $b_1$ -- $b_2$ correlation
explicitly and showed for the first time that it indeed explains the
observed behavior of $Q$ for equilateral triangles calculated from 
SDSS galaxies in linear regimes.

The major findings of the present paper are summarized as 
follows. 
\begin{itemize}
\item Even with the presence of redshift distortion and complicated
survey shape effects, $b_1$ can be accurately estimated in $k<0.2h{\rm
Mpc}^{-1}$. So $b_1$ estimated from SDSS galaxies is expected to reflect
the true value in real space.
\item The values of $Q_b$ for equilateral triangles from simulated halos, 
mock galaxies based on HOD model, and SDSS galaxies do not show any 
clear dependence on $b_1$. The independence on $b_1$ is not an 
artifact from the redshift-space 
distortion contamination, but is a consequence of the intrinsic 
correlation between $b_2/b_1$ and $b_1$.
\item The HOD model, equation (\ref{eq:HOD}), whose parameters are 
determined so as to reproduce the observed two-point statistics alone, 
seems to be also successful in predicting the value of $b_2$.
\end{itemize}

There are a few remaining tasks which should be done following the
present result.  First, it is interesting to compare the current
analysis in Fourier space with that in configuration space.  As long as
nonlinear effect is negligible, $Q_m$ and $Q_b$ in k-space are expected
to be the same as those in configuration space.  Nevertheless we would
like to make sure of it, and to further explore the connection between
bispectra and the three-point correlation functions of halos and
galaxies. 
Second, the success of HOD to reproduce the correlation between 
$b_2/b_1$ and $b_1$ at the current level is very encouraging. 
Naturally it is important to further improve the HOD model. 
Third, the present paper considers only equilateral configuration 
of Fourier space triangles for $Q$. In reality, however, the different 
shape of triangles has a wealth of information.
While the $b_1$ dependence of $Q$ barely vanishes for equilateral 
triangles due to the correlation between $b_1$ and $b_2/b_1$, this is 
not the case for arbitrary triangle shapes. Therefore it is interesting 
to explore if the $b_1$ dependence of $Q$ for the other triangle shapes 
and to see if the results are consistent with the present analysis and/or 
to constrain higher order biasing coefficients more tightly.
The work along the line is now in progress.

\bigskip

We thank Akihito Shirata and Kohji Yoshikawa for useful discussions.
The N-body simulations were performed at ADAC (Astronomical Data
Analysis Center) of the National Astronomical Observatory, Japan.
K.~Y. and C.~H. acknowledge the support from Grants-in-Aid for 
Japan Society for the Promotion of Science Fellows. This research 
was partly supported by Grant-in-Aid for Scientific Research of Japan 
Society for Promotion of Science (No.16340053, 17740139).

Funding for the SDSS and SDSS-II has been provided by the Alfred P.
Sloan Foundation, the Participating Institutions, the National Science
Foundation, the U.S. Department of Energy, the National Aeronautics and
Space Administration, the Japanese Monbukagakusho, the Max Planck
Society, and the Higher Education Funding Council for England.  The SDSS
Web Site is http://www.sdss.org/.

The SDSS is managed by the Astrophysical Research Consortium for the
Participating Institutions. The Participating Institutions are the
American Museum of Natural History, Astrophysical Institute Potsdam,
University of Basel, Cambridge University, Case Western Reserve
University, University of Chicago, Drexel University, Fermilab, the
Institute for Advanced Study, the Japan Participation Group, Johns
Hopkins University, the Joint Institute for Nuclear Astrophysics, the
Kavli Institute for Particle Astrophysics and Cosmology, the Korean
Scientist Group, the Chinese Academy of Sciences (LAMOST), Los Alamos
National Laboratory, the Max-Planck-Institute for Astronomy (MPIA), the
Max-Planck-Institute for Astrophysics (MPA), New Mexico State
University, Ohio State University, University of Pittsburgh, University
of Portsmouth, Princeton University, the United States Naval
Observatory, and the University of Washington.

\bigskip

\end{document}